\documentclass[11pt,showpacs,preprintnumbers,superscriptaddress,amsmath,amssymb,nofootinbib]{revtex4}

\usepackage{graphicx}
\usepackage{dcolumn}
\usepackage{bm}
\usepackage{amssymb}
\usepackage{amsmath}
\usepackage{epsfig}    
\usepackage{color}
\usepackage{slashed}
\usepackage{ulem} 
\usepackage[hypertex]{hyperref}
\begin{document} 

\title{Unitarity bound in the most general two Higgs doublet model}
\preprint{UT-HET 107}

\author{Shinya Kanemura}
\email{kanemu@sci.u-toyama.ac.jp}
\affiliation{Department of Physics, University of Toyama, \\3190 Gofuku, Toyama 930-8555, Japan}
\author{Kei Yagyu}
\email{K.Yagyu@soton.ac.uk}
\affiliation{School of Physics and Astronomy, University of Southampton, Southampton, SO17 1BJ, United Kingdom}

\begin{abstract}

We investigate unitarity bounds in the most general two Higgs doublet model without a discrete $Z_2$ symmetry nor CP conservation. 
S-wave amplitudes for two-body elastic scatterings of Nambu-Goldstone bosons and physical Higgs bosons 
are calculated at high energies for all possible initial and final states (14 neutral, 8 singly-charged and 3 doubly-charged states). 
We obtain analytic formulae for the block-diagonalized scattering matrix 
by the classification of the two body scattering states using the conserved quantum numbers at high energies. 
Imposing the condition of perturbative unitarity to the eigenvalues of the scattering matrix, 
constraints on the model parameters can be obtained. 
We apply our results to constrain the
mass range of the next--to--lightest Higgs state in the model. 

\end{abstract}
\maketitle

\section{Introduction}

The Higgs boson was discovered at LHC in 2012~\cite{LHC_ATLAS,LHC_CMS}, and its mass and coupling constants 
turned out to be consistent with the predictions in the standard model (SM)~\cite{LHC_ATLAS2,LHC_CMS2}. 
However, although the Higgs boson was found, 
the structure of the Higgs sector remains unknown. 
There are possibilities of non-minimal Higgs sectors which contain 
additional Higgs bosons. 
They also can satisfy the current data. 
In fact, in large number of new physics scenarios, extended Higgs sectors are predicted. 
The structure of the Higgs sector is strongly connected to these new physics scenarios. 
Therefore, the Higgs sector is a probe of new physics. 

Although parameters of an extended Higgs sector are basically free, 
they can be constrained by imposing some compelling theoretical conditions, such as 
those of perturbative unitarity, vacuum stability and triviality. 
The requirement of perturbative unitarity~\cite{lqt} is known to give a conservative bound on the parameters of a model, 
beyond which the perturbation calculation does not work. 
It goes without saying that the parameters can also be constrained by taking into account various 
experimental results such as electroweak precision data~\cite{LEP}, observables in flavour physics~\cite{Flavour}, 
the Higgs boson signal strength~\cite{LHC_ATLAS2,LHC_CMS2} and so on. 
By the combination of the bounds
from the theoretical consistency and those from the experimental data, a parameter space in extended Higgs sectors can be further restricted.

The theoretical constraints have been intensively studied in the two Higgs doublet model (THDM). 
Nice reviews for the THDMs have been presented in Refs.~\cite{HHG,Rev}.  
Previously, the unitarity bound was mainly 
studied for the model with a (softly-broken) discrete $Z_2$ symmetry~\cite{GW} with CP-conservation~\cite{Huffel,Maalampi,KKT,Akeroyd,Das}.
The softly-broken discrete $Z_2$ symmetry is phenomenologically important in order to avoid flavor changing neutral currents (FCNCs)\footnote{As an alternative way
to avoid the tree level FCNCs, the aligned THDMs~\cite{Pich} have also been known, where two Yukawa matrices are assumed to be proportional with other.}, 
under which 
there are four types of Yukawa interactions~\cite{Barger,Grossman,Akeroyd-Yukawa,AKTY,Su,Logan}. 
Bounds from vacuum stability and triviality have also been studied in Refs.~\cite{Ma,Sher1,Sher2,Kasai} and in Refs.~\cite{Inoue,Kasai,Dey}, respectively, in $Z_2$ symmetric THDMs.  
These theoretical bounds have been used to limit magnitudes of cross sections and decay rates of tree level processes~\cite{TreeProcess,AKTY,ktyy}, 
radiative corrections due to additional Higgs bosons to the Higgs couplings~\cite{KOSY,Coupling} and one-loop induced processes~\cite{Loop-Induced}.  

However, analyses in the most general THDM without the $Z_2$ symmetry is getting important 
as an effective description of more various new physics scenarios, such as supersymmetric SMs with non-holomorphic Yukawa couplings~\cite{Babu-Kolda} 
and also general models with CP-violation~\cite{CPV} which is required for successful scenario of electroweak baryogenesis~\cite{ewbg1,ewbg2,ewbg3}. 

In this letter, we investigate unitarity bounds in the most general THDM without the $Z_2$ symmetry nor CP-conservation. 
We calculate S-wave amplitudes for two-body elastic scatterings of Nambu-Goldstone (NG) bosons and physical scalar bosons 
at high energies for all possible initial and final states (14 neutral~\cite{KKT}, 8 singly-charged~\cite{Akeroyd} and 3 doubly-charged states). 
By choosing appropriate bases of the scattering states which are deduced 
using the conserved quantum numbers at high energies such as the hypercharge, weak isospin and the third component of the latter~\cite{Ginzburg_CPV}, 
the scattering matrix is given as a block-diagonalized form with at most $4\times 4$ submatrices. 
Thus, all of the eigenvalues of the scattering matrix can be easily evaluated numerically. 
The analytic result for the block diagonalized S-wave matrix is consistent with that given in Ref.~\cite{Ginzburg}. 
By requiring that each of the eigenvalues is not too large to break validity of perturbation calculation, 
the model parameters, e.g., the masses of extra Higgs bosons and mixing angles, can be constrained. 
Our results can be useful to constrain parameter spaces whenever one evaluates physics quantities in the most general THDM. 

We then numerically demonstrate that the bound on the mass $M_{\text{2nd}}$ of the second lightest Higgs boson, whatever it is, is obtained 
by inputting the mass of the discovered Higgs boson $h$ under the assumption of 
non-zero deviations in the Higgs boson coupling $hVV$ with the weak boson $(V=W$ and $Z)$ from the SM value. 
Currently, the $hVV$ coupling was measured with $\sim 10\%$ accuracy by the LHC Run-I experiment (see Ref.~\cite{10p} and references therein), and that
is expected to be measured more accurately at future collider experiments.
If a deviation is detected in the $hVV$ coupling, we can obtain the constraint on the region of $M_{\text{2nd}}$ even without its direct discovery.

\section{ Model setup }
\subsection{Higgs potential}

The most general Higgs potential under the $SU(2)_L\times U(1)_Y$ gauge symmetry is given by
\begin{align}
V&=m_1^2|\Phi_1|^2+m_2^2|\Phi_2|^2-(m_3^2\Phi_1^\dagger \Phi_2 +\text{h.c.})\notag\\
& +\frac{1}{2}\lambda_1|\Phi_1|^4+\frac{1}{2}\lambda_2|\Phi_2|^4+\lambda_3|\Phi_1|^2|\Phi_2|^2+\lambda_4|\Phi_1^\dagger\Phi_2|^2
 +\frac{1}{2}\left[\lambda_5(\Phi_1^\dagger\Phi_2)^2+\text{h.c.}\right] \notag\\
&+\left[\lambda_6|\Phi_1|^2\Phi_1^\dagger \Phi_2+\lambda_7|\Phi_2|^2\Phi_1^\dagger\Phi_2+\text{h.c.}\right], \label{pot_thdm1}
\end{align}
where $\Phi_i$ ($i=1,2$) are the isospin doublet scalar fields with hypercharge $Y=1/2$. 
In general, 
$m_1^2$, $m_2^2$ and $\lambda_1$-$\lambda_4$ are real, while $m_3^2$ and $\lambda_5$-$\lambda_7$ are complex. 
Thus, there are totally fourteen real parameters. 
If there is a symmetry in the potential, the number of parameters is reduced. 
For example, when the potential is exact (softly-broken) $Z_2$ invariant under the transformation of $(\Phi_1,\Phi_2) \to (\Phi_1, -\Phi_2)$, 
the $m_3^2$, $\lambda_6$ and $\lambda_7$ ($\lambda_6$ and $\lambda_7$) terms are forbidden. 

By using the $U(1)_Y$ invariance and rephasing the doublet fields, the vacuum expectation values (VEVs) of the two doublet fields can be 
taken to be real without loss of generality~\cite{Maria,Osland,Santos}. 
The two doublet fields are then described in terms of the component fields as 
\begin{align}
\Phi_i=\left[\begin{array}{c}
\omega_i^+\\
\frac{1}{\sqrt{2}}(v_i+h_i+iz_i)
\end{array}\right],\hspace{3mm}(i=1,2), \label{Eq:parametrizations}
\end{align}
where $v_1$ and $v_2$ are the VEVs of $\Phi_1$ and $\Phi_2$, respectively, 
which are satisfied $v=\sqrt{v_1^2+v_2^2}=(\sqrt{2}G_F)^{-1/2}\simeq 246$ GeV.
By introducing $\tan\beta=v_2/v_1$, two VEVs are described by $v$ and $\tan\beta$ as the usual notation. 
In the following, both the VEVs are assumed to be non-zero, except for 
the case of the inert doublet model discussed in Appendix. 

The stationary conditions of the scalar potential are given as follows
\begin{align}
\left. \frac{\partial V}{\partial \varphi_a}\right|_0 = 0,\quad (\varphi_a = h_1,~h_2,~z_1,~\text{and}~z_2), \label{stationaly}
\end{align}
where the left hand side of the above equation for each $\varphi$ is calculated by 
\begin{align}
&\left.\frac{\partial V}{\partial h_1}\right|_0=
vc_\beta^{}\left[m_1^2
-M^2s^2_\beta 
+\frac{v^2}{2}\left(\lambda_1c^2_\beta
+\lambda_{345}s^2_\beta
+3\lambda_6^Rs_\beta^{}c_\beta^{}
+\lambda_7^Rs^2_\beta\tan\beta  \right) \right], \label{tadpole1}\\
&\left.\frac{\partial V}{\partial h_2}\right|_0
=vs_\beta^{}\left[m_2^2
-M^2\cos^2\beta
+\frac{v^2}{2}\left(\lambda_2s^2_\beta
+\lambda_{345}c^2_\beta+\lambda_6^Rc^2_\beta\cot\beta 
+3\lambda_7^Rs_\beta^{}c_\beta^{} \right)\right], \label{tadpole2}\\
&\left.\frac{\partial V}{\partial z_1}\right|_0
=-v\tan\beta\left.\frac{\partial V}{\partial z_2}\right|_0
=vs_\beta^{}\left[\text{Im}\, m_3^2
+\frac{v^2}{2}\left(\lambda_5^I s_\beta^{}c_\beta^{}
+\lambda_6^Ic^2_\beta
+\lambda_7^Is^2_\beta \right)\right], \label{tadpole3}
\end{align}
where we used the abbreviation of $s_\theta^{}=\sin \theta$ and $c_\theta^{}=\cos \theta$. 
In Eqs.~(\ref{tadpole1})-(\ref{tadpole3}), we introduced 
\begin{align}
\lambda_{345}= \lambda_3+\lambda_4 +\lambda_5^R, \quad
M^2=\frac{\text{Re}\, m_3^2}{s_\beta c_\beta}, \label{bigm}
\end{align}
and 
\begin{align}
\lambda_{k}^R = \text{Re}\,\lambda_{k},\quad 
\lambda_{k}^I = \text{Im}\,\lambda_{k},\quad (k=5,~6,~7). 
\end{align}
From the stationary conditions in Eq.~(\ref{stationaly}), 
we can eliminate $m_1^2$, $m_2^2$ and Im\,$m_3^2$ in the Higgs potential.   

In order to calculate the masses for the scalar bosons, it is convenient to introduce the so-called Higgs basis~\cite{HB} defined as
\begin{align}
\left(\begin{array}{c}
\Phi_1\\
\Phi_2
\end{array}\right)=
\left(\begin{array}{cc}
c_\beta & -s_\beta\\
s_\beta & c_\beta
\end{array}\right)
\left(\begin{array}{c}
\Phi\\
\Psi
\end{array}\right), 
\end{align}
where 
\begin{align}
\Phi=\left[
\begin{array}{c}
G^+\\
\frac{1}{\sqrt{2}}(v+h_1'+ iG^0)
\end{array}\right],\quad
\Psi=\left[
\begin{array}{c}
H^+\\
\frac{1}{\sqrt{2}}(h_2'+ih_3')
\end{array}\right], \label{Higgs-basis}
\end{align}
where $G^\pm$ and $G^0$ are the NG bosons which are absorbed into the longitudinal components of $W^\pm$ and $Z$ by the Higgs mechanism, respectively.  
The physical singly-charged scalar state is denoted as $H^\pm$. 
The three neutral states $h_1'$, $h_2'$ and $h_3'$ are not the mass eigenstates at this stage, which generally mix with each other. 

\subsection{Mass spectrum}

We give the mass formulae for the Higgs bosons in the most general case. 
First, the squared mass of $H^\pm$ is calculated by
\begin{align}
m_{H^\pm}^2&=M^2-\frac{v^2}{2}\left(\lambda_4+\lambda_5^R -\lambda_6^R \cot\beta -\lambda_7^R \tan\beta\right). \label{mchsq}
\end{align}

Next, the mass term for the neutral scalar states is expressed by the $3\times 3$ matrix as  
\begin{align}
V_{\text{neutral}}^{\text{mass}}&=\frac{1}{2}(h_1',h_2',h_3')
\left(\begin{array}{ccc}
M_{11}^2 & M_{12}^2 & M_{13}^2\\
M_{12}^2 &M_{22}^2  & M_{23}^2\\
M_{13}^2 &M_{23}^2  & M_{33}^2
\end{array}\right)
\left(\begin{array}{c}
h_1' \\
h_2' \\
h_3' 
\end{array}\right),   \label{massmat}
\end{align}
where each of the matrix elements is given by 
\begin{subequations}
\begin{align}
M_{11}^2&=v^2\left[\lambda_1c^4_\beta+\lambda_2 s^4_\beta+\frac{\lambda_{345}}{2}s^2_{2\beta}+ 2(\lambda_6^Rc^2_\beta + \lambda_7^Rs^2_\beta )s_{2\beta} \right], \label{m11sq} \\
M_{22}^2&=M^2+\frac{v^2}{4}\Big[
(\lambda_1+\lambda_2-2\lambda_{345})s^2_{2\beta}  -2\lambda_6^R(s_{4\beta}^{}+\cot\beta) -2\lambda_7^R(s_{4\beta}^{}-\tan\beta)\Big]  ,\label{m22sq}\\
M_{33}^2&
= M^2-v^2\left(\lambda_5^R + \frac{\lambda_6^R }{2}\cot\beta + \frac{\lambda_7^R}{2}\tan\beta\right), \label{m33sq} \\
M_{12}^2&=\frac{v^2}{2}\left[(\lambda_2 s^2_\beta-\lambda_1c^2_\beta + \lambda_{345}c_{2\beta}^{} )s_{2\beta}^{}
+2\lambda^R_6(2c_{2\beta}^{} -1)c^2_\beta + 2\lambda_7^R(2c_{2\beta}^{} +1)s^2_\beta \right] , \label{m12sq} \\
M_{13}^2&
= -\frac{v^2}{2}\left(\lambda_5^I s_{2\beta}^{} + 2\lambda_6^I c^2_\beta  + 2\lambda_7^I s^2_\beta \right), \\
M_{23}^2&
= -\frac{v^2}{2}\left(\lambda_5^Ic_{2\beta}^{} - \lambda_6^Ic^2_\beta  + \lambda_7^Is^2_\beta \right). 
\end{align}
\label{mateven}
\end{subequations}
The mass eigenvalues $m_{H_i}^2$ ($i=1$-3) are obtained by introducing 
the $3\times 3$ orthogonal matrix $R$ as 
\begin{align}
R^T_{ik}M_{kl}^2R_{lj} = \text{diag}(m_{H_1}^2,m_{H_2}^2,m_{H_3}^2), 
\end{align}
where $m_{H_1}\leq m_{H_2}\leq m_{H_3}$ is assumed. 
Mass eigenstates for the neutral Higgs bosons are also defined using $R$ as 
\begin{align}
\left(\begin{array}{c}
h_1' \\
h_2' \\
h_3' 
\end{array}\right)=
R
\left(\begin{array}{c}
H_1 \\
H_2 \\
H_3 
\end{array}\right), \label{neut}
\end{align}
where $H_1$ is defined to be the SM-like Higgs boson with the mass of about 125 GeV, i.e., $m_{H_1}^{}\simeq 125$ GeV. 
In the following, we represent $H_1$ and its mass $m_{H_1}^{}$ by $h$ and $m_h^{}$, respectively.  
In the CP-conserving limit, $H_2$ and $H_3$ respectively correspond to the additional CP-even ($H$)
and CP-odd ($A$) Higgs bosons. 
The mass formulae for the CP-conserving case will be discussed in the next subsection. 
We here note that $R$ can be described by three mixing angles~\cite{Osland}. 
In our numerical analysis given in Sec.~IV, 
the matrix elements of $R$ are derived by inputting the mass matrix elements given in Eq.~(\ref{mateven}).   

By using Eqs.~(\ref{mchsq}), (\ref{m11sq}), (\ref{m22sq}), (\ref{m22sq}), and (\ref{m33sq}), 
the five parameters ($\lambda_{1\text{-}4}$ and $\lambda_5^R$) can be rewritten as follows
\begin{align}
\lambda_1v^2 &= M_{11}^2 + (M_{22}^2-M^2)\tan^2\beta -2M_{12}^2\tan\beta
 +\frac{1}{2}(\lambda_7^R\tan^2\beta-3\lambda_6^R )\tan\beta, \label{lam1}\\
%%%
\lambda_2v^2 &= M_{11}^2 + (M_{22}^2-M^2)\cot^2\beta +2M_{12}^2\cot\beta
+\frac{1}{2}(\lambda_6^R \cot^2\beta - 3\lambda_7^R)\cot\beta, \\
%%%
\lambda_3v^2 &= M_{11}^2-(M_{22}^2+M^2)+2M_{12}^2\cot2\beta +2m_{H^\pm}^2
              -\frac{1}{2}(\lambda_6^R \cot\beta  +\lambda_7^R \tan\beta ), \\
\lambda_4v^2&=M^2+M_{33}^2-2m_{H^\pm}^2-\frac{1}{2}(\lambda_6^R\cot\beta + \lambda_7^R\tan\beta  ),\\
\lambda_5^Rv^2&=M^2-M_{33}^2-\frac{1}{2}(\lambda_6^R\cot\beta + \lambda_7^R\tan\beta  ). \label{physical}
\end{align}  
Therefore, we can choose thirteen input parameters as $v$, $\tan\beta$, $M^2$, $M_{11}^2$, $M_{22}^2$, $M_{12}^2$, $M_{33}^2$, $m_{H^\pm}^{2}$, $\lambda_{6,7}^R$ and $\lambda_{5,6,7}^I$. 
Instead of $M_{11}^2$, $M_{22}^2$ and $M_{12}^2$, we can take $\tilde{\alpha}$, $\tilde{m}_H^2$ and $\tilde{m}_h^2$ as inputs by
\begin{align}
M_{11}^2 &= \tilde{m}_h^2s^2_{\beta-\tilde{\alpha}} + \tilde{m}_H^2c^2_{\beta-\tilde{\alpha}}, \label{m11}\\
M_{22}^2 &= \tilde{m}_h^2c^2_{\beta-\tilde{\alpha}} + \tilde{m}_H^2s^2_{\beta-\tilde{\alpha}}, \label{m22} \\
M_{12}^2 &= (\tilde{m}_h^2 - \tilde{m}_H^2)s_{\beta-\tilde{\alpha}}^{}c_{\beta-\tilde{\alpha}}^{}.  \label{m12} 
\end{align}
In addition, we rewrite $ M_{33}^2$ as 
\begin{align}
M_{33}^2 = \tilde{m}_A^2.  
\end{align}
In the CP-conserving limit ($\text{Im}\,m_3^2\to 0$ and $\lambda_{5,6,7}^I \to 0$), 
$\tilde{m}_{h}^2$, $\tilde{m}_{H}^2$, $\tilde{m}_{A}^2$ and $\tilde{\alpha}$ describe the physical squared masses for 
$h$, $H$ and $A$, and the mixing angle for the CP-even Higgs state, respectively. 

Consequently, by using experimental values of $v$ ($=246$ GeV) and $m_h$ ($=125$ GeV), 
eleven input parameters can be chosen as follows
\begin{align}
\{\tilde{m}_{H}^{2},~\tilde{m}_A^2,~m_{H^\pm}^2,~M^2,~\tan\beta,~\sin(\beta-\tilde{\alpha}),~|\lambda_{6}|,~|\lambda_{7}|,~\theta_{5},~\theta_{6},~\theta_{7}  \}, \label{input1}
\end{align}
where $\theta_{5,6,7}$ are the complex phases of $\lambda_{5,6,7}$. 
Notice that $\tilde{m}_h$ is determined so as to keep $m_h=125$ GeV.

\section{Unitarity bounds}

We calculate the S-wave amplitude matrix for the elastic $BB'\to B''B'''$ scatterings in the high energy limit, where
the fields $B$, $B'$, $B''$ and $B'''$ represent either $W_L^\pm$, $Z_L$, $H_1$, $H_2$, $H_3$ or $H^\pm$. 
In this case, all the longitudinal components of 
the weak gauge boson states can be replaced by the corresponding NG boson states because of the equivalence theorem. 
Furthermore, only the scalar boson contact interactions contribute to the S-wave amplitude. 
Therefore, the calculation of the S-wave amplitude matrix is quite simply done just by extracting the coefficient of the scalar boson quartic terms. 

There are 14 neutral, 8 singly-charged and 3 doubly-charged 2 body scalar boson channels.  
First, the 14 neutral channels are expressed in the weak eigenbasis as 
\begin{align}
|\omega_i^+\omega_i^-\rangle,~
\frac{1}{\sqrt{2}}|z_iz_i\rangle,~
\frac{1}{\sqrt{2}}|h_ih_i \rangle,~
|h_iz_i\rangle,~
|\omega_1^+\omega_2^-\rangle,~
|\omega_2^+\omega_1^-\rangle,~
|z_1 z_2 \rangle,~
|h_1 h_2\rangle,~
|h_1 z_2 \rangle,~
|h_2 z_1\rangle, 
\end{align}
where $i=1,2$. 
By taking an appropriate basis transformation, we obtain the following block-diagonalized S-wave amplitude matrix:
\begin{align}
a_0^0 = 
\frac{1}{16\pi}
\left(
\begin{array}{cccc}
X_{4\times 4} & 0 & 0 & 0 \\
0 & Y_{4\times 4} & 0 & 0 \\
0 & 0 & Z_{3 \times 3} & 0 \\
0 & 0 & 0             &Z_{3\times 3} 
\end{array}\right), \label{block}
\end{align}
where each submatrix is given by
\begin{align}
&X_{4\times 4}=\begin{pmatrix}
3\lambda_1 & 2\lambda_3 + \lambda_4 & 3\sqrt{2}\lambda_6^R & 3\sqrt{2}\lambda_6^I \\
2\lambda_3 + \lambda_4 &3\lambda_2&   3\sqrt{2}\lambda_7^R    & 3\sqrt{2}\lambda_7^I \\
3\sqrt{2}\lambda_6^R & 3\sqrt{2}\lambda_7^R           & \lambda_3+2\lambda_4+3\lambda_5^R  & 3\lambda_5^I \\
3\sqrt{2}\lambda_6^I & 3\sqrt{2}\lambda_7^I           &  3\lambda_5^I  & \lambda_3+2\lambda_4-3\lambda_5^R \\
\end{pmatrix}, \\
&Y_{4\times 4} =\begin{pmatrix}
\lambda_1 & \lambda_4 & \sqrt{2}\lambda_6^R& \sqrt{2}\lambda_6^I \\
\lambda_4 & \lambda_2 & \sqrt{2}\lambda_7^R& \sqrt{2}\lambda_7^I \\
\sqrt{2}\lambda_6^R & \sqrt{2}\lambda_7^R& \lambda_3+\lambda_5^R & \lambda_5^I \\
\sqrt{2}\lambda_6^I & \sqrt{2}\lambda_7^I& \lambda_5^I & \lambda_3-\lambda_5^R
\end{pmatrix}, \\
&Z_{3\times 3}=\begin{pmatrix}
\lambda_1 & \lambda_5^R + i\lambda_5^I & \sqrt{2}(\lambda_6^R + i\lambda_6^I) \\
\lambda_5^R - i\lambda_5^I & \lambda_2 & \sqrt{2}(\lambda_7^R - i\lambda_7^I)   \\
\sqrt{2}(\lambda_6^R -i\lambda_6^I)    & \sqrt{2}(\lambda_7^R + i\lambda_7^I) & \lambda_3+\lambda_4
\end{pmatrix}.  
\end{align}
Each of the submatrices is obtained in the following basis:
\begin{align}
\Psi_i^N(X_{4\times 4}) = 
& \frac{1}{\sqrt{2}}\left|\omega_1^+\omega_1^- +\frac{z_1^{}z_1^{}}{2} + \frac{h_1^{}h_1^{}}{2}\right\rangle,~
\frac{1}{\sqrt{2}}\left|\omega_2^+\omega_2^- +\frac{z_2^{}z_2^{}}{2} + \frac{h_2^{}h_2^{}}{2}\right\rangle ,~\notag\\
&\frac{1}{2}\left|\omega_1^+\omega_2^- + \omega_2^+\omega_1^- + z_1^{}z_2^{} + h_1^{}h_2^{}   \right\rangle,~ 
\frac{1}{2}\left|-i\omega_1^+\omega_2^- + i\omega_2^+\omega_1^- - h_1^{}z_2^{} + h_2^{}z_1^{}   \right\rangle ,  \\
\Psi_j^N(Y_{4\times 4})=& \frac{1}{\sqrt{2}}\left|\omega_1^+\omega_1^- -\frac{z_1^{}z_1^{}}{2} - \frac{h_1^{}h_1^{}}{2}\right\rangle,~
\frac{1}{\sqrt{2}}\left|\omega_2^+\omega_2^- - \frac{z_2^{}z_2^{}}{2} - \frac{h_2^{}h_2^{}}{2}\right\rangle ,~\notag\\
&\frac{1}{2}\left|-\omega_1^+\omega_2^- - \omega_2^+\omega_1^- + z_1^{}z_2^{} + h_1^{}h_2^{}   \right\rangle,~ 
\frac{1}{2}\left|i\omega_1^+\omega_2^- - i\omega_2^+\omega_1^- - h_1^{}z_2^{} + h_2^{}z_1^{}   \right\rangle ,  \\
\Psi_k^N(Z_{3\times 3})=& \frac{1}{\sqrt{2}}\left|+ \frac{z_1^{}z_1^{}}{2} - \frac{h_1^{}h_1^{}}{2} + ih_1^{}z_1^{} \right\rangle,~
\frac{1}{\sqrt{2}}\left|+ \frac{z_2^{}z_2^{}}{2} - \frac{h_2^{}h_2^{}}{2} + ih_2^{}z_2^{} \right\rangle,~ \notag\\
&\frac{1}{2}\left|+ z_1^{}z_2^{} - h_1^{}h_2^{} +ih_1z_2^{}+ih_2z_1^{}  \right\rangle, \\
\Psi_l^{N'}(Z_{3\times 3})=& \frac{1}{\sqrt{2}}\left|- \frac{z_1^{}z_1^{}}{2} + \frac{h_1^{}h_1^{}}{2} + ih_1^{}z_1^{} \right\rangle,~
\frac{1}{\sqrt{2}}\left|- \frac{z_2^{}z_2^{}}{2} + \frac{h_2^{}h_2^{}}{2} + ih_2^{}z_2^{} \right\rangle,~ \notag\\
&\frac{1}{2}\left|- z_1^{}z_2^{} + h_1^{}h_2^{} +ih_1z_2^{}+ih_2z_1^{}  \right\rangle, 
\end{align}
where the indices $i$ and $j$ ($k$ and $l$) run over 1-4 (1-3). 
Each of the above bases give the submatrix indicated in the parenthesis. 
We note that at the high energy limit, 
the hypercharge $Y$, the isospin $I$ and the third component of the isospin $I_3$ are used to classify 
the two body scattering states~\cite{Ginzburg_CPV}, namely $\Phi_a\times \Phi_b$ with $Y=1$ and $\Phi_a\times \tilde{\Phi}_b$ 
($\tilde{\Phi}_b\equiv i\tau_2 \Phi_b^*$) with $Y=0$. 
In fact, the above bases are obtained by finding the two scalar states 
which belong to the same set of the quantum numbers, i.e.,    
the states in the $\Psi_i^N(X_{4\times 4})$, 
$\Psi_j^N(Y_{4\times 4})$, 
$\Psi_k^N(Z_{4\times 4})$ and 
$\Psi_l^{N'}(Z_{4\times 4})$ bases respectively belong to the state with ($Y$, $I$, $I_3)=(0,0,0)$, $(0,1,0)$, $(1,1,-1)$ and $(1,1,-1)$. 

Next, eight (positive) singly-charged channels are expressed as:
\begin{align}
|\omega_i^+z_i\rangle,~
|\omega_i^+h_i\rangle,~
|\omega_1^+z_2\rangle,~
|\omega_1^+h_2\rangle,~
|\omega_2^+z_1\rangle,~
|\omega_2^+h_1\rangle. 
\end{align}
We obtain the following block diagonalized S-wave amplitude matrix:
\begin{align}
a_0^+ = 
\frac{1}{16\pi}\left(
\begin{array}{cccc}
Y_{4\times 4} & 0 & 0  \\
0 & Z_{3\times 3} & 0  \\
0 & 0 & \lambda_3-\lambda_4  
\end{array}\right).  \label{block2}
\end{align}
Each of the submatrices and the eigenvalue $\lambda_3-\lambda_4$ are obtained in the following basis:
\begin{align}
\Psi_i^C(Y_{4\times 4})=& \frac{1}{\sqrt{2}}\left|-i\omega_1^+z_1^{} + \omega_1^+h_1 \right\rangle,~
\frac{1}{\sqrt{2}}\left|-i\omega_2^+z_2^{} + \omega_2^+h_2 \right\rangle, \notag\\
&\frac{1}{2}\left|-i\omega_1^+z_2^{} - i\omega_2^+z_1^{} + \omega_1^+ h_2 + \omega_2^+ h_1  \right\rangle,~
\frac{1}{2}\left|-\omega_1^+z_2^{} + \omega_2^+z_1^{} -i \omega_1^+ h_2 +i \omega_2^+ h_1  \right\rangle, \notag\\
\Psi_j^C(Z_{3\times 3})=& \frac{1}{\sqrt{2}}\left|i\omega_1^+z_1^{} + \omega_1^+h_1  \right\rangle,~
\frac{1}{\sqrt{2}}\left|i\omega_2^+z_2^{} + \omega_2^+h_2    \right\rangle ,~
\frac{1}{2}\left|i\omega_1^+z_2^{} + i\omega_2^+z_1^{} + \omega_1^+ h_2 + \omega_2^+ h_1   \right\rangle,~  \notag\\
\Psi^C(\lambda_3-\lambda_4)=&\frac{1}{2}\left|-i\omega_1^+z_2^{} + i\omega_2^+z_1^{} - \omega_1^+ h_2 + \omega_2^+ h_1   \right\rangle, 
\end{align}
where $i=1$-4 and $j=1$-3. 
Similar to the discussion for the neutral states, 
the states in the bases $\Psi_i^C(Y_{4\times 4})$, $\Psi_i^C(Z_{3\times 3})$ and $\Psi^C(\lambda_3-\lambda_4)$
respectively correspond to the states with 
($Y$, $I$, $I_3)=(0,1,1)$, $(1,1,0)$ and $(1,0,0)$.

Finally, three (positive) doubly-charged channels are expressed as:
\begin{align}
\frac{1}{\sqrt{2}}|\omega_1^+\omega_1^+\rangle, ~~ 
\frac{1}{\sqrt{2}}|\omega_2^+\omega_2^+\rangle, ~~
|\omega_1^+\omega_2^+\rangle. 
\end{align}
We obtain 
\begin{align}
a_0^{++} = 
\frac{1}{16\pi}Z_{3\times 3}. 
\end{align}
We note that the negative charged states are obtained by taking charge conjugation for each positive charged channel, which give the same set of eigenvalues of the matrix 
for the corresponding positive charged channels. 

When we impose symmetries in the potential, we obtain the block-diagonalized S-wave amplitude matrix with a smaller size of submatrices. 
For example, if we assume the CP-invariance (for the case of $\lambda_5^I=\lambda_6^I=\lambda_7^I=0$), the maximal size of the submatrices reduces into $3\times 3$, 
or if we impose the (softly-broken) $Z_2$ symmetry (for the case of $\lambda_6=\lambda_7=0$), 
the maximal size of the submatrices reduces into $2\times 2$ as they have already known in Refs.~\cite{KKT,Akeroyd}.

In order to constrain the parameters in the potential, we impose the following condition for each eigenvalue of 
the S-wave amplitude matrix: 
\begin{align}
|\text{Re}(x_i)| < \xi,~~i=1,\dots 12,  \label{lqtbound}
\end{align}
where $\xi$ is conventionally taken to be 1/2~\cite{HHG} or 1~\cite{lqt}, and 
\begin{align}
x_i = \{\text{Eigenvalues}(X_{4\times 4}),~\text{Eigenvalues}(Y_{4\times 4}),~\text{Eigenvalues}(Z_{3\times 3}),~\lambda_3-\lambda_4 \},  \label{eigen} 
\end{align}
with all $x_i$ being real due to the hermitian nature of the S-wave amplitude matrix. 
We call the bound given by the inequality (\ref{lqtbound}) as the unitarity bound. 
We note that 
the analytic expressions for the eigenvalues in Eq.~(\ref{eigen}) are given by 
solving the fourth and third order equations. 
In general, the solutions of such an equation is given as a too complicated form to explicitly show in this letter. 
Therefore, we do not show the explicit formula, and we numerically calculate the eigenvalues in the numerical study given in the next section.

\section{Numerical Studies}

We here discuss the constraint on the parameter space using the unitarity bound.  
The unitarity bound in Eq.~(\ref{lqtbound}) sets upper limits on $x_i$ ($i=1,\dots,~12$) given in Eq.~(\ref{eigen}), which are expressed by combinations of $\lambda$ parameters. 
Using Eqs.~(\ref{lam1})-(\ref{m12}), this constraint can be translated into the bound on physical parameters such as the masses of Higgs bosons and mixing angles.

It is worthful to mention here that we can obtain an upper limit on the masses of extra Higgs bosons even when the SM-like Higgs boson $h~(=H_1)$ 
coupling with the gauge bosons ($hVV$, $V=W,Z$) slightly deviates from the SM prediction. 
This can be intuitively understood as follows. 

First, in the Higgs basis defined in Eq.~(\ref{Higgs-basis}), the kinetic terms for the Higgs fields are given by 
\begin{align}
{\cal L}_{\text{kin}} = |D_\mu \Phi|^2 + |D_\mu \Psi|^2, \label{lkin}
\end{align}
where $D_\mu$ is the covariant derivative. 
The Higgs-Gauge-Gauge type vertex, i.e., $h_1'VV$, only comes from the first term of Eq.~(\ref{lkin}) at the tree level. 
In the mass eigenbasis, 
the ratio of the $hVV$ coupling $g_{hVV}^{}$ in the THDM to that of SM is then expressed 
by using the rotation matrix $R$ given in Eq.~(\ref{neut}) as 
\begin{align}
\kappa_V^{} \equiv \frac{g_{hVV}^{\text{THDM}}}{g_{hVV}^{\text{SM}}} = R_{11}. 
\end{align}
We note that $R_{11}$ corresponds to $\sin(\beta-\alpha)$ in the CP-conserving case as it is seen in Eq.~(\ref{sinba}). 
Thus, the non-zero deviation in the $hVV$ couplings from the SM prediction comes from the mixing effect of neutral Higgs bosons, i.e., $R_{11}\neq 1$. 

Second, when there is no mixing among the neutral Higgs bosons, the mass of $h$ and 
those of the extra Higgs bosons $H_2$ and $H_3$ are schematically expressed as $\lambda_i v^2$ and $M^2 + \lambda_j v^2$, respectively, as 
it is seen in Eqs.~(\ref{m11sq}), (\ref{m22sq}) and (\ref{m33sq}). 
Therefore, in the no mixing case, the upper bound on the masses of extra Higgs bosons cannot be obtained, because
they can be taken to be as large as possible by using the $M^2$ dependence. 
In other words, we can take the decoupling limit~\cite{Gunion} of the extra Higgs bosons by taking $M^2 \gg v^2$.  

On the other hand, if there is non-zero mixing among the neutral Higgs bosons, i.e., $\kappa_V^{}\neq 1$, 
$M^2$ dependence appears in $m_{h}^2$ which must be kept to be about (125 GeV)$^2$. 
Therefore, we cannot take a too large value of $M^2$ in that case, because we need a large cancellation of the $M^2$ 
contribution to $m_{h}^2$ by the $\lambda_i v^2$ term which must be excluded by the unitarity bound. 
Therefore, we can obtain an upper limit on the masses of extra Higgs bosons as long as the $hVV$ coupling is deviated from the SM prediction. 

\begin{figure}[t]
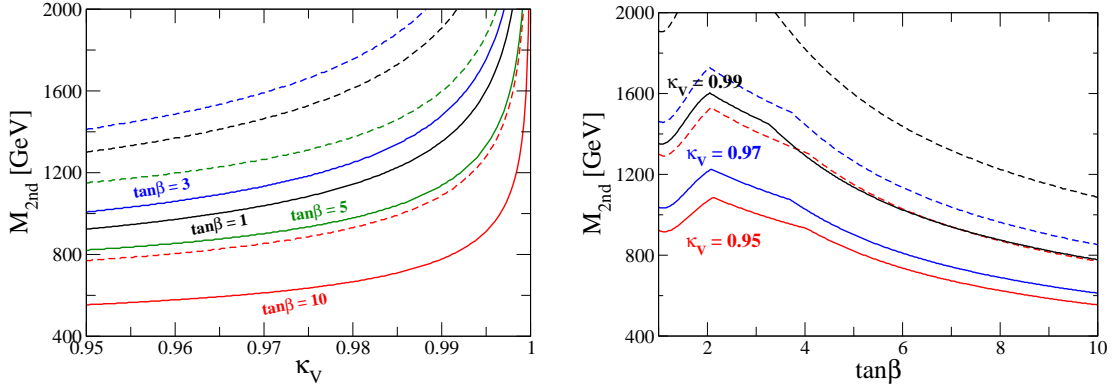

\begin{center}
\includegraphics[width=70mm]{uni_z2_v2.eps}\hspace{5mm}
\includegraphics[width=70mm]{uni_z2_del_v2.eps}
\caption{The upper limit on the mass of the second lightest Higgs boson $M_{\text{2nd}}$ from the unitarity bound in the softly-broken $Z_2$ symmetric case. 
The left and right panels show the $\kappa_V^{}$ and $\tan\beta$ dependences of the upper limit 
for fixed values of $\tan\beta=1$ (black), 3 (blue), 5 (green) and 10 (red) and 
those of $\kappa_V^{}=0.99$ (black), 0.97 (blue) and 0.95 (red), respectively. 
The solid (dashed) curve shows the case with $\xi=1/2$ (1) in Eq.~(\ref{lqtbound}). 
In both panels, we take $m_{H^\pm}^{}=m_A^{}=m_H^{}\,(=M_{\text{2nd}})$, and scan the value of $M^2$.    }
\label{uni1}
\end{center}
\end{figure}

\begin{figure}[t]
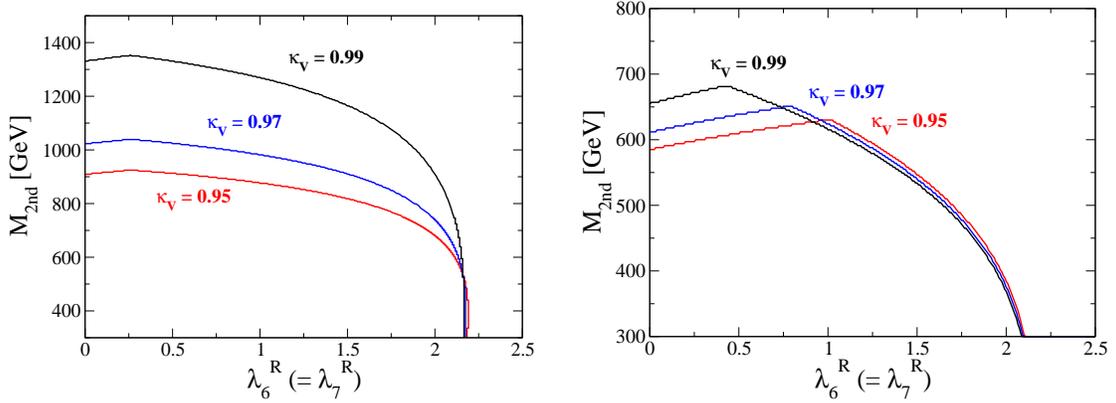

\begin{center}
\includegraphics[width=70mm]{fig1.eps}\hspace{5mm}
\includegraphics[width=70mm]{fig1b.eps}
\caption{
The upper limit on the mass of the second lightest Higgs boson $M_{\text{2nd}}$ from the unitarity bound as a function of  $\lambda_6^R~(=\lambda_7^R)$ 
in the case without the CP-violating phases. 
The black, blue and red curves show the case of $\kappa_V^{}=0.99$, 0.97 and 0.95, respectively. 
We take $m_{H^\pm}^{}=m_A^{}=m_H^{}\,(=M_{\text{2nd}})$ and $\tan\beta=1$.  
The left (right) panel shows the case of $M=m_{H^\pm}^{}$ ($M=m_{H^\pm}^{}/2$). }
\label{uni2}
\end{center}
\end{figure}

In the following, we numerically show the bound on the mass of the second lightest Higgs boson denoted as $M_{\text{2nd}}$ 
by fixing $m_{h}^{}=125$ GeV and $\kappa_V^{}$.  
We assume that the SM-like Higgs boson $h$ is the lightest of all the Higgs bosons, so that 
$M_{\text{2nd}}$ is defined by 
\begin{align}
M_{\text{2nd}} &\equiv \text{Min}(m_{H^\pm},~m_{A}^{},~m_{H}),~\text{for the CP-conserving case}, \notag\\
M_{\text{2nd}} &\equiv \text{Min}(m_{H^\pm},~m_{H_2}^{},~m_{H_3}),~\text{for the CP-violating case}.
\end{align}

We first consider the softly-broken $Z_2$ symmetric and CP-conserving case, i.e., $\lambda_{6,7}=0$ and $\theta_{5}=0$. 
In this case, we have six free parameters $m_{H^\pm}^{}$, $m_A^{}$, $m_H^{}$, $M^2$, $\tan\beta$ and $\sin(\beta-\alpha)\,(=\kappa_V^{})$. 
In Fig.~\ref{uni1}, we show the upper limit on $M_{\text{2nd}}$ as a function of $\kappa_V^{}$ (left panel) 
and $\tan\beta$ (right panel) in the case of $m_{H^\pm}^{}=m_A^{}=m_H^{}\,(=M_{\text{2nd}})$. 
In the left (right) panel, each of the curves show the cases of $\tan\beta$ ($\kappa_V^{}$) to be 1, 3, 5 and 10 (0.99, 0.97 and 0.95). 
The value of $M^2$ is scanned in the both figures.
The solid (dashed) curves show the case with $\xi=1/2$ (1) in Eq.~(\ref{lqtbound}) for the comparison
of the strength of the unitarity constraint with the two cases.  Hereafter, we take $\xi=1/2$. 
From these figures, we can see that the stronger limit on $M_{\text{2nd}}$ is obtained when $1-\kappa_V^{}$ is taken to be larger values. 
Our results for this case are consistent with those given in Ref.~\cite{ktyy}.

\begin{figure}[t]
\begin{center}
\includegraphics[width=70mm]{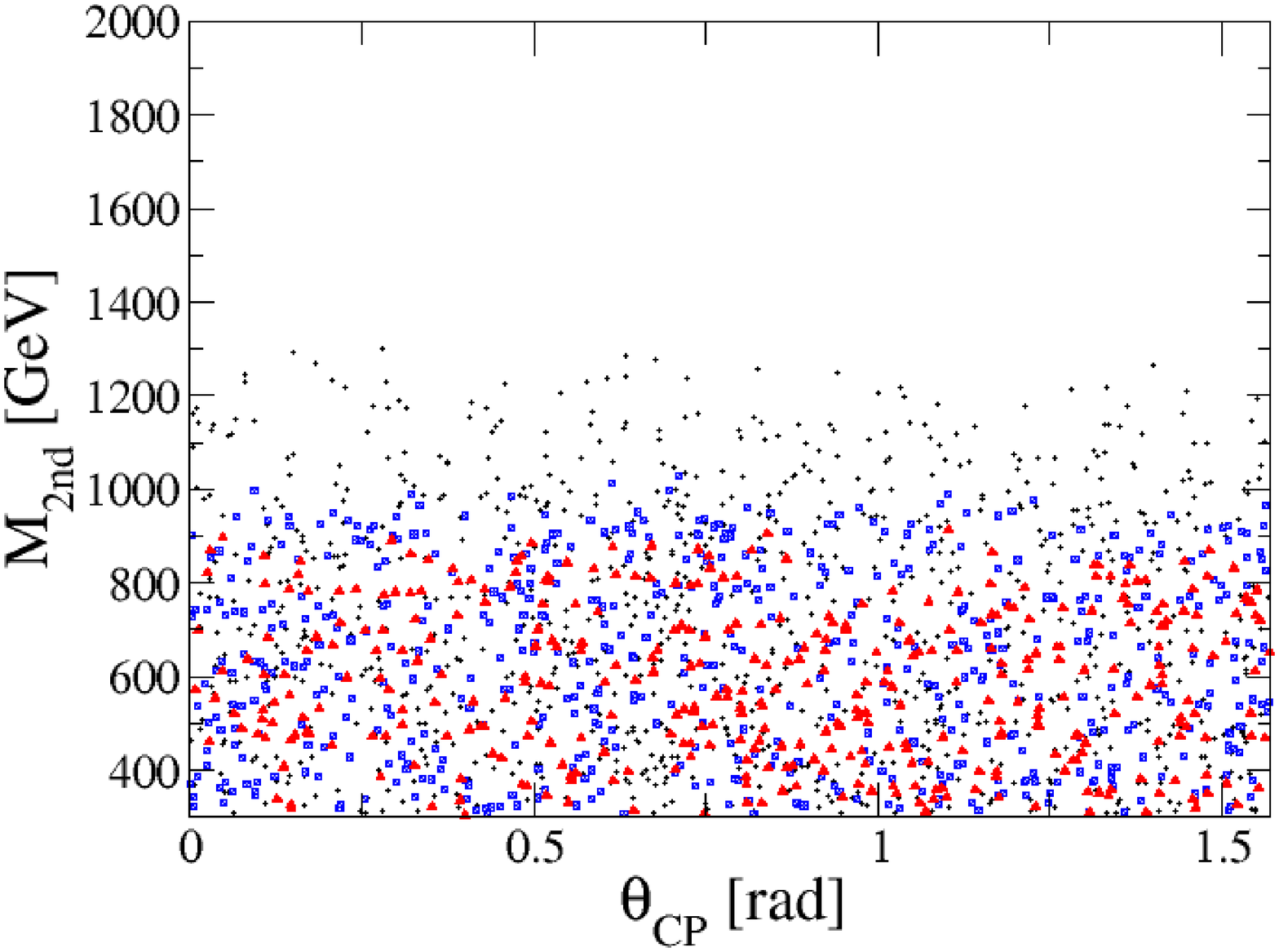}\hspace{5mm}
\includegraphics[width=70mm]{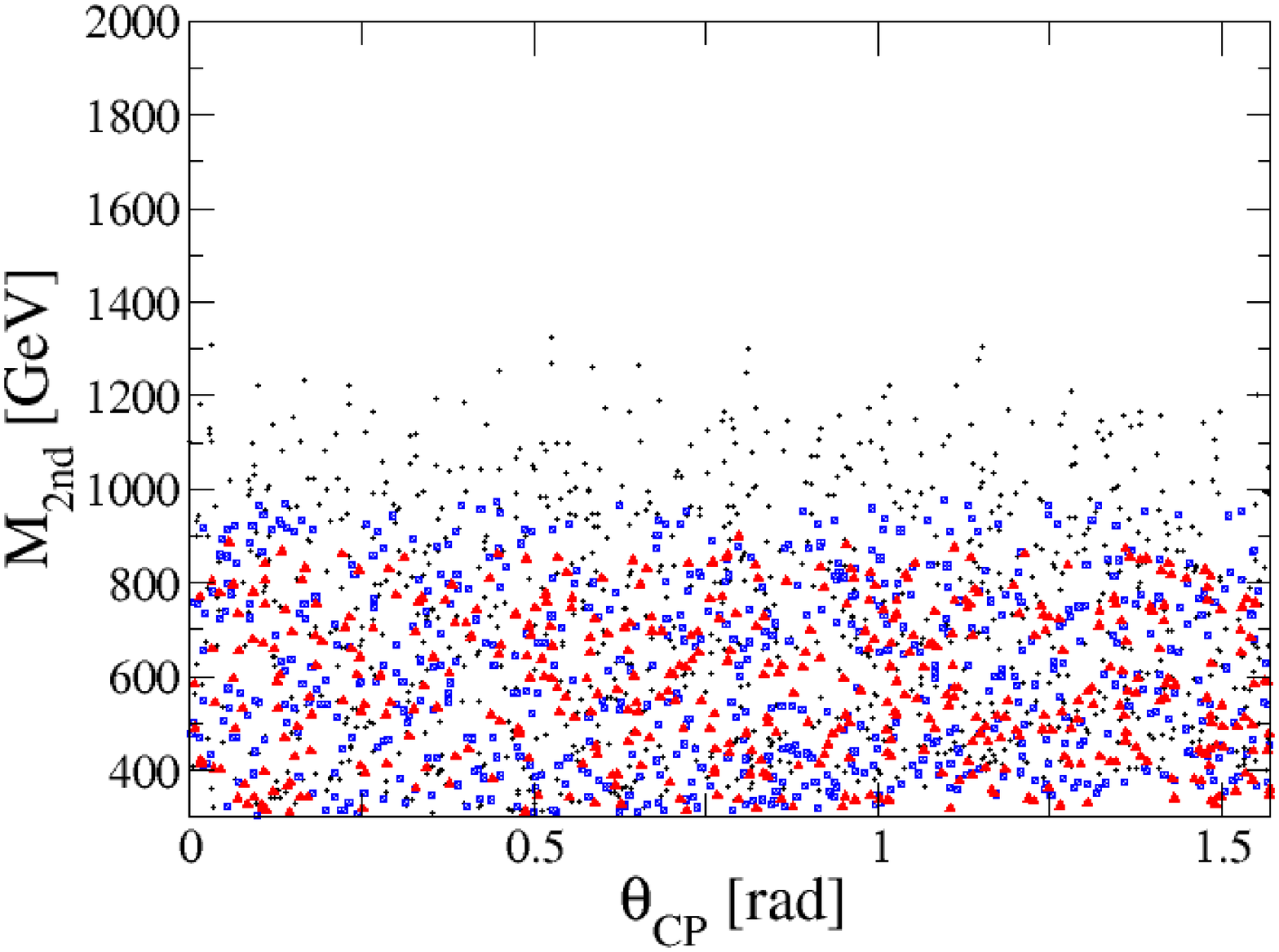}\\ \vspace{10mm}
\includegraphics[width=70mm]{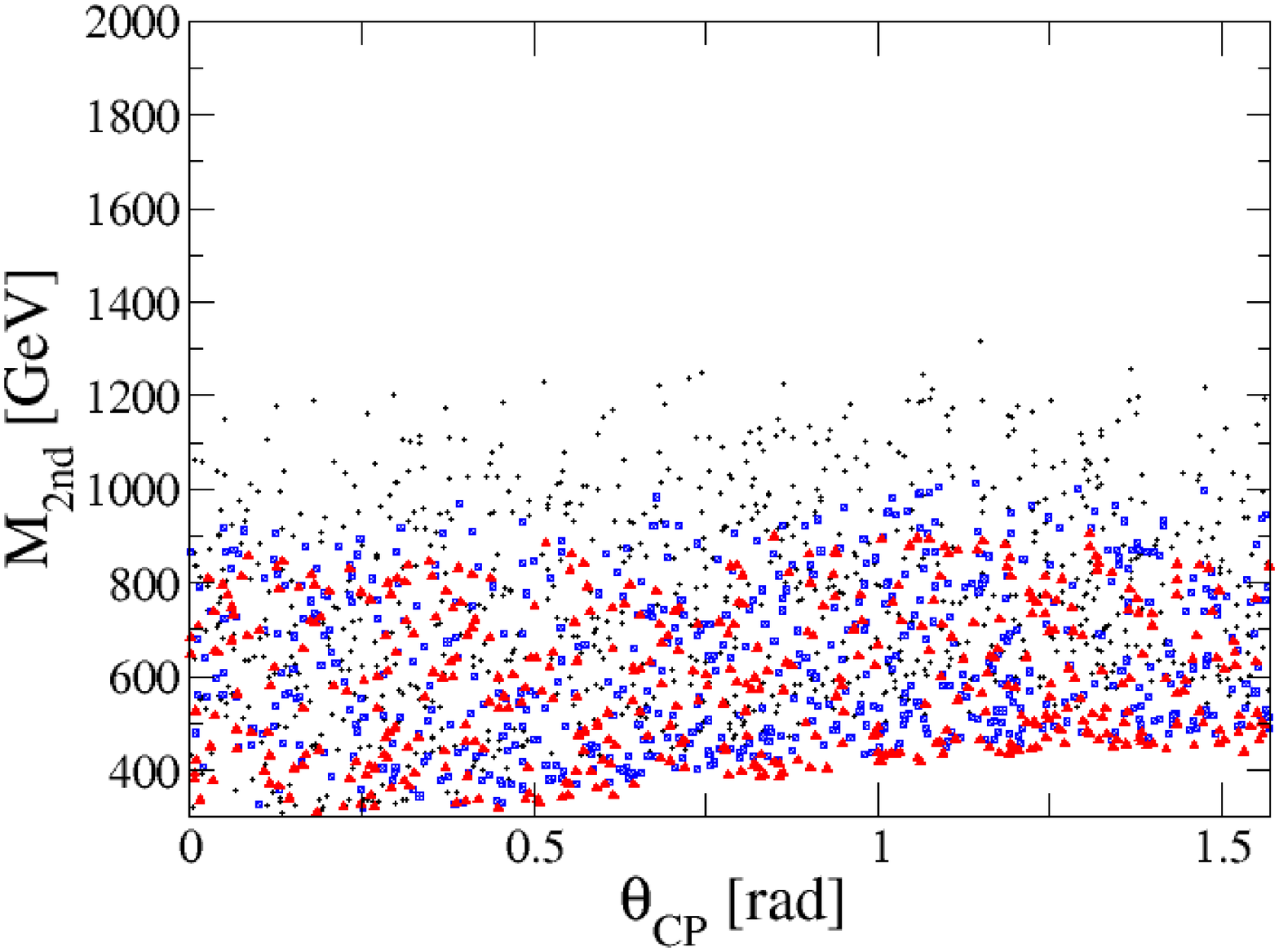}\hspace{5mm}
\includegraphics[width=70mm]{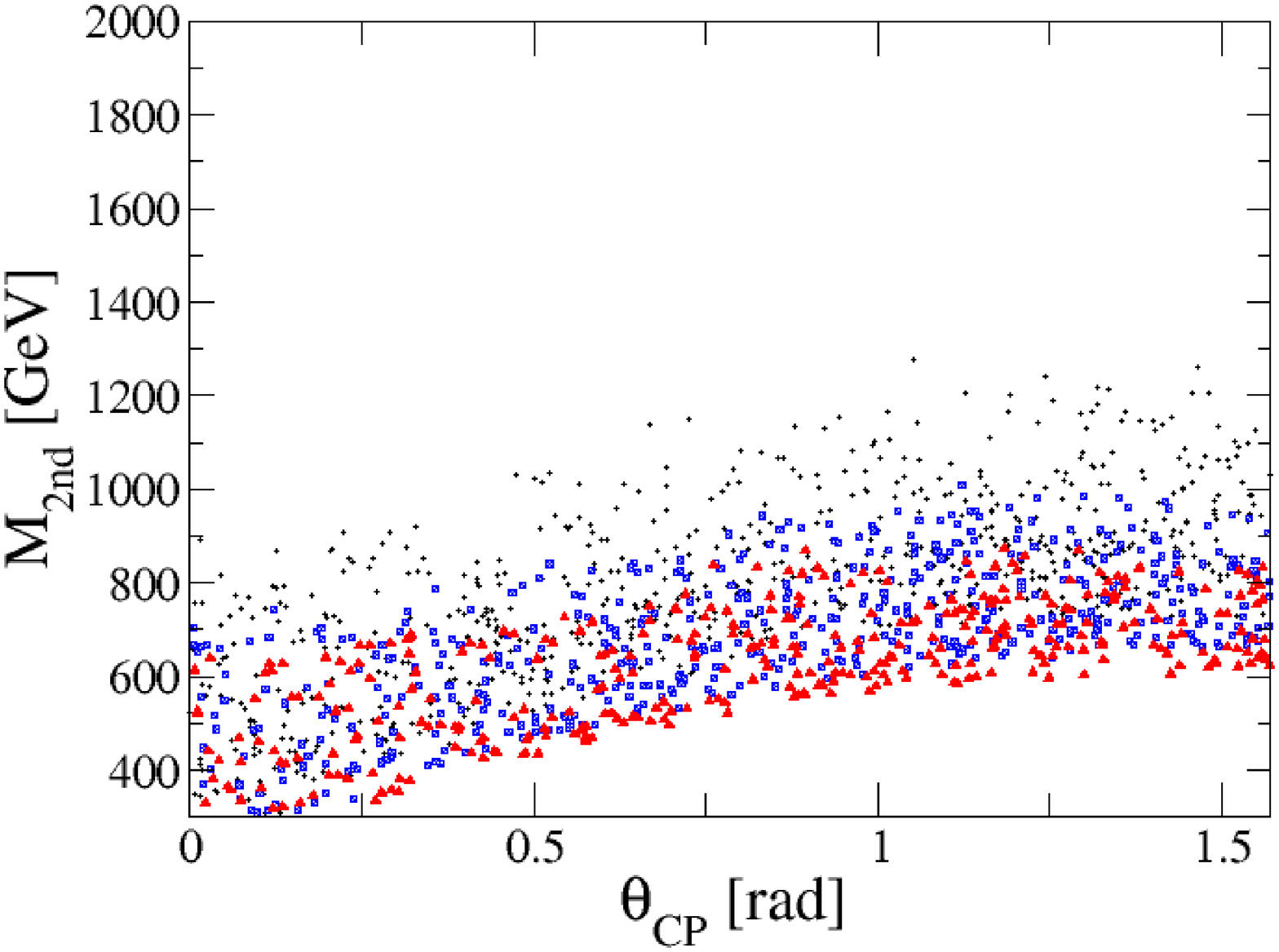}
\caption{Scatter plots for the allowed parameter space from the unitarity bound on the $\theta_{\text{CP}}(=\theta_6=\theta_7)$ and $M_{\text{2nd}}$ plane in the case of 
$|\lambda_6|(=|\lambda_7|)=0.2$ (upper-left), 0.5 (upper-right), 1 (lower-left) and 2 (lower-right). 
For all the plots, we take $m_{H^\pm}^2=\tilde{m}_{A}^{2}=\tilde{m}_H^{2}=M^2$, $\theta_5=0$ and $\tan\beta=1$, 
and scan the $m_{H^\pm}^2$ and $\sin(\beta-\tilde{\alpha})$ parameters.
The black (circle), blue (square) and red (triangle) dots show the allowed points for the cases of 
$\kappa_V^{}=0.98\pm 0.01$, $0.96\pm 0.01$ and $0.94\pm 0.01$, respectively.  }
\label{uni3}
\end{center}
\end{figure}

Next, we consider the case without the softly-broken $Z_2$ symmetry but with CP-conservation, i.e., $\theta_{5,6,7}=0$. 
In Fig.~\ref{uni2}, we show the upper limit on $M_{\text{2nd}}$ 
as a function of $\lambda_6^R\,(=\lambda_7^R)$ in the case of $\tan\beta = 1$ and $m_{H^\pm}=m_{A}^{}=m_H^{}\,(=M_{\text{2nd}})$. 
The value of $M^2$ is taken to be $M^2=M_{\text{2nd}}^2$ (left panel) and $M^2=(M_{\text{2nd}}/2)^2$ (right panel). 
In both the panels, the black, blue and red curves show the cases with $\kappa_V^{}=0.99$, 0.97 and 0.95, respectively. 
Clearly, the stronger bound is obtained in the case of $M^2=(M_{\text{2nd}}/2)^2$ (right panel) than the case of $M^2 = M_{\text{2nd}}^2$, 
where the appearance of kink at around $\lambda_6^R=0.5$, 0.8 and 1 
is due to an interchange of eigenvalues which break the condition of Eq.~(\ref{lqtbound}). 
When $\lambda_6^R$ is taken to be larger than about 2.2, there is no solution to satisfy the unitarity bound.  
We confirm that the maximally allowed value of $M_{\text{2nd}}$ is obtained in the case of $M^2 \simeq M_{\text{2nd}}^2$. 

Finally, we consider the most general case with CP-violation. 
In this case, we have eleven input parameters with $v=246$ GeV and $m_h=125$ GeV as shown in (\ref{input1}). 
In order to satisfy $m_h=125$ GeV, we scan the $\tilde{m}_h$ parameter for each fixed value of the other parameters.  
In Fig.~\ref{uni3}, we show the allowed parameter regions from the unitarity bound on the $\theta_{\text{CP}}(=\theta_6=\theta_7)$ 
and $M_{\text{2nd}}$ plane in the case of $\tan\beta =1$, $m_{H^\pm}^2 = \tilde{m}_A^2=\tilde{m}_{H}^2 = M^2$ and $\theta_5=0$. 
The upper-left, upper-right, lower-left and lower-right panels respectively show the cases with   
$|\lambda_6|=|\lambda_7|=0.2$, 0.5, 1 and 2. 
In all the panels, the circle (black), square (blue) and triangle (red) points satisfy the unitarity bound and $\kappa_V^{}=0.98\pm 0.01$, $0.96\pm 0.01$ and 
$0.94\pm 0.01$, respectively. 
The values of $m_{H^\pm}$ and $\sin(\beta-\tilde{\alpha})$ are scanned. 
We find that, in addition to the upper limit on $M_{\text{2nd}}$, there is the lower limit especially in the case with $\theta_{\text{CP}}\neq 0$. 
The lower bound becomes higher when we take a larger value of $|\lambda_{6}|~(=|\lambda_7|)$. 
The appearance of the lower limit can be understood in the following way. 
If we take a non-zero value of $\theta_{\text{CP}}$, it gives a non-zero value of the off-diagonal mass matrix elements $M_{13}^2$ and $M_{23}^2$ (see Eq.~(\ref{mateven})), 
which gives non-zero mixings and/or mass splittings among the neutral Higgs bosons. 
On the other hand, we now fix the value of $\kappa_V^{}$, so that it restricts the possible amount of the mixing, and it also requires 
a non-zero mass splitting among the neutral Higgs bosons. 
Because the SM-like Higgs boson $h$ which we suppose the lightest of all has the mass of 125 GeV, the non-zero mass splitting turns out to be the lower limit on $M_{\text{2nd}}$. 

We here comment on the constraint from electric dipole moments (EDMs) on the parameter space in the THDM. 
As it has been well known that if a model has an additional source of the CP-violation, its magnitude is constrained by EDMs. 
In Refs.~\cite{EDM1,EDM2}, the constraint on parameter space from EDMs has been investigated in the softly-broken $Z_2$ symmetric THDMs. 
In the THDMs, EDMs constrain the allowed region of $\theta_{\text{CP}}$ depending on the parameter set. 
Recent study on
the collider phenomenology of the CP-violating THDM is found in Ref.~\cite{Rui}, 
in which the constraint on the complex parameter such as $\lambda_5$ 
from EDMs and the unitarity bound turns out to be important.

\section{Conclusions}

We have investigated unitarity bounds in the most general two Higgs doublet model without 
a discrete $Z_2$ symmetry nor CP conservation. 
We have computed the S-wave amplitudes for two-body elastic scatterings of the NG bosons and physical Higgs bosons 
at high energies for all possible initial and final states.  
By choosing the appropriate bases, the scattering amplitude matrix is given to be  
the block-diagonalized form, and thus the eigenvalues can be easily evaluated numerically. 
We have constrained the parameter space of the model by using the unitarity bound.  
By fixing the mass of the discovered Higgs boson $h$ to be 125 GeV and assuming a non-zero deviation in the $hVV$ couplings from 
the SM values, 
there is an upper limit on the mass $M_{\text{2nd}}$ of the second lightest Higgs boson. 
Therefore, by using the precisely measured $hVV$ couplings at future collider experiments, 
we can constrain the allowed region of $M_{\text{2nd}}$ if
a deviation of the $hVV$ coupling from the SM limit is found even without its direct discovery. 
Our results can be useful to constrain parameter spaces whenever one evaluates physics quantities in the most general THDM. 

\section*{Acknowledgments}

This work was supported, in part, by Grant-in-Aid for Scientific Research No. 23104006  (SK)
and Grand H2020-MSCA-RISE-2014 no. 645722 (Non Minimal Higgs) (SK).
This work was also supported by a JSPS postdoctoral fellowships for research abroad (KY).

\begin{appendix}

\section{Mass spectrum in specific cases }

We present the mass formulae of the scalar bosons for specific cases of the THDM, i.e., 
the case with CP-conservation but without the $Z_2$ symmetry, and 
that with the unbroken $Z_2$ symmetry usually referred as the inert doublet model.  

\subsection{The most general case without CP-violating phases}

When we take the limit of $\lambda_{5,6,7}^I \to 0$, 
the CP symmetry is restored in the Higgs potential, 
and the potential is then described by the ten parameters. 
The mass of $H^\pm$ is unchanged and is given in Eq.~(\ref{mchsq}). 

For the mass matrix of neutral Higgs bosons, 
the $3\times 3$ matrix given in Eq.~(\ref{massmat}) becomes 
a block-diagonalized form, i.e., the $2\times 2$ plus $1\times 1$ form. 
The first $2\times 2$ part with the basis of ($h_1'$,$h_2'$) corresponds to the mass matrix for the CP-even Higgs states, 
and the remained $1\times 1$ with the basis of $h_3'\,(\equiv A)$ does the squared mass of the CP-odd Higgs boson. 
For the CP-even states, we obtain the mass eigenstates by introducing the mixing angle $\beta-\alpha$ by
\begin{align}
\left(\begin{array}{c}
h_1' \\
h_2'
\end{array}\right)=
\left(
\begin{array}{cc}
s_{\beta-\alpha} & c_{\beta-\alpha} \\
c_{\beta-\alpha} & -s_{\beta-\alpha}
\end{array} \right)
\left(\begin{array}{c}
h\\
H
\end{array}\right), \label{sinba}
\end{align}
where $h$ is defined as the SM-like Higgs boson. 

The squared masses of $A$ is given by 
\begin{align}
&m_A^2= M_{33}^2. 
\end{align}
Masses for $H$ and $h$ and the mixing angle $\beta-\alpha$ are respectively given by solving Eqs.~(\ref{m11}), (\ref{m22}) and (\ref{m12}) 
with the replacement $(\tilde{\alpha}, \tilde{m}_H^2, \tilde{m}_h^2)\to (\alpha,m_H^2,m_h^2)$:
\begin{align}
&m_H^2=M_{11}^2c^2_{\beta-\alpha}+M_{22}^2s^2_{\beta-\alpha}+2M_{12}^2s_{\beta-\alpha}^{}c_{\beta-\alpha}^{}, \label{mHsq}  \\
&m_h^2=M_{11}^2s^2_{\beta-\alpha}+M_{22}^2 c^2_{\beta-\alpha}-2M_{12}^2s_{\beta-\alpha}^{}c_{\beta-\alpha-}^{}, \label{mhsq} \\
&\tan 2(\beta-\alpha)=\frac{2M_{12}^2}{M_{22}^2-M_{11}^2}.
\end{align} 

Consequently, eight input parameters can be chosen as follows
\begin{align}
\{m_{H}^{2},~m_A^2,~m_{H^\pm}^2,~M^2,~\tan\beta,~\sin(\beta-\alpha),~\lambda_{6}^R,~\lambda_{7}^R  \},  \label{input2}
\end{align}
and experimental values of $v$ and $m_h$. 
In this parameter choice, the five parameters $\lambda_{1\text{-}4}$ and $\lambda_{5}^R$ are derived by Eqs.~(\ref{lam1})-(\ref{m12}) 
with the replacement $(\tilde{\alpha}, \tilde{m}_H^2, \tilde{m}_h^2)\to (\alpha,m_H^2,m_h^2)$.

The mass formulae in the well-known softly-broken $Z_2$ symmetric case with the CP-invariance are easily obtained by taking 
$\lambda_6^R = \lambda_7^R =0$, which agree with the formulae given, e.g., in Ref.~\cite{KOSY}. 

\subsection{The inert doublet case}

When we impose an exact $Z_2$ symmetry in the potential, where the two doublets are transformed as 
$\Phi_1 \to +\Phi_1$ and $\Phi_2 \to -\Phi_2$, the potential is given for the case with $m_3^2=\lambda_6=\lambda_7=0$ in Eq.~(\ref{pot_thdm1}). 
In this case, the phase of $\lambda_5$ can be dropped without loss of generality, so that CP-violation does not occur. 
The VEV of $\Phi_2$ must be assumed to be zero to avoid the spontaneous breakdown of the $Z_2$ symmetry, otherwise 
the $m_{3}^2\Phi_1^\dagger \Phi_2+\text{h.c.}$ terms are generated through the $\lambda$ terms, and  
the inert nature, i.e., without couplings to SM particles, is then lost via the mixing with the active Higgs field, namely, the SM-like Higgs boson.

The squared masses of the scalar bosons are calculated by 
\begin{align}
&m_{h}^2 = \lambda_1 v^2,\quad
m_H^2 = m_2^2  +  \frac{v^2}{2}\lambda_{345} ,\quad
m_A^2 = m_2^2    +  \frac{v^2}{2} (\lambda_3+\lambda_4-\lambda_5^R),\quad
m_{H^\pm}^2 = m_2^2 +\frac{v^2}{2}   \lambda_3. 
\end{align}

\end{appendix}

\end{document}